\documentclass[ aps, showpacs, showkeys, nofootinbib, floatfix, superscriptaddress]{revtex4}

\usepackage{amsfonts}

\usepackage{amssymb}
\usepackage{amsmath}
\usepackage{graphicx}
\usepackage{hyperref}

\begin{document}

\title{Combined constraints on modified Chaplygin gas model from cosmological observed data: Markov Chain Monte Carlo approach}

 \author{Jianbo Lu}
 \email{lvjianbo819@163.com}
 \affiliation{Department of Physics, Liaoning Normal University, Dalian 116029, P. R. China}
 \author{Lixin Xu}
 \email{lxxu@dlut.edu.cn}
 \affiliation{School of Physics and Optoelectronic Technology, Dalian University of Technology, Dalian, 116024, P. R. China}
 \author{Yabo Wu \footnote{Corresponding author: ybwu61@163.com}}
 \affiliation{Department of Physics, Liaoning Normal University, Dalian 116029, P. R. China}
 \author{Molin Liu}
 \affiliation{College of Physics and Electronic Engineering, Xinyang Normal University, Xinyang, 464000, P. R. China}

\begin{abstract} We use the Markov Chain Monte Carlo method to  investigate a global
constraints on  the modified Chaplygin gas (MCG) model as the
unification of dark matter and dark energy from  the latest
observational data: the Union2 dataset of type supernovae Ia (SNIa),
the observational Hubble data (OHD),  the cluster X-ray gas mass
fraction, the baryon acoustic oscillation (BAO), and the
 cosmic microwave background (CMB) data. In a
 flat universe,  the constraint results for MCG model are,
 $\Omega_{b}h^{2}=0.02263^{+0.00184}_{-0.00162}$ ($1\sigma$) $^{+0.00213}_{-0.00195}$ $(2\sigma)$,
 $B_{s}=0.7788^{+0.0736}_{-0.0723}$ ($1\sigma$) $^{+0.0918}_{-0.0904}$ $(2\sigma)$,
 $\alpha=0.1079^{+0.3397}_{-0.2539}$ ($1\sigma$) $^{+0.4678}_{-0.2911}$ $(2\sigma)$,
  $B=0.00189^{+0.00583}_{-0.00756}$ ($1\sigma$) $^{+0.00660}_{-0.00915}$ $(2\sigma)$,
 and $H_{0}=70.711^{+4.188}_{-3.142}$ ($1\sigma$) $^{+5.281}_{-4.149}$ $(2\sigma)$.
\end{abstract}


\maketitle

\section{$\text{Introduction}$}

{Recently, mounting cosmic observations suggest that the expansion
of present universe is speeding up rather than slowing down
\cite{SNeCMBLSS}.  And it indicates that baryon matter component is
about 5\% of the  total energy density, and about 95\%  of the
energy density in the universe is invisible, including dark matter
(DM) and dark energy (DE). In addition, it is shown that DE takes up
about two-thirds of the total energy density from cosmic
observations. In theory many kinds of DE models \cite{DEmodels} have
already been constructed to explore the DE properties.

 Chaplygin gas (CG) and its  generalized model have been widely
studied for interpreting the accelerating universe
\cite{GCG}\cite{GCGpapers}. The CG model can be obtained from the
string Nambu-Goto action in the light cone coordinate
\cite{CG-string}. For generalized Chaplygin gas (GCG), it emerges as
a effective fluid of a generalized d-brane in a $(d+1, 1)$ space
time, and its action can be written as a generalized Born-Infeld
form \cite{GCG-action}.
  Considering that the application of string theory in principle is
in very high energy when the quantum effects is important in early
universe \cite{CG-string}, the quantum cosmological studis of the CG
and the GCG has been well investigated in Ref. \cite{CG-string} and
\cite{GCG-quantum}. In addition, one knows that the most attractive
 property for these models is, two unknown dark sections
in universe--dark energy and dark matter can be unified by using an
exotic equation of state. It is worthwhile to study the unified
models of dark sections for other  generalization of CG.

A simple and popular generalization  relative to the GCG model is
that it is extended to a  form by adding a barotropic term, referred
to as the modified Chaplygin gas (MCG) \cite{MCG}.
  The correspondences between the MCG and the ordinary scalar field \cite{MCG}, the  tachyon theory \cite{tachyon,MCG}, and
  the holographic dark energy density \cite{MCG-hde} of the universe  have been
studied.  In principle, MCG  represents the evolution of the
universe starting from the radiation era to the era dominated by the
cosmological constant \cite{MCG-epoch,MCG-radiation,MCG-sound}. In
addition, MCG is also applied to inflation theory
\cite{MCG-inflation}. For the more discussion on MCG model, please
see Refs. \cite{MCG-interaction,MCG-constraint,MCG-otherpapers}.
 In this paper, we use the Markov Chain
Monte Carlo (MCMC) technique to constrain this more general unified
candidate, MCG. The used observational data include, the Union2 data
of type Ia supernovae (SNIa) \cite{557Union2}, the observational
Hubble data (OHD) \cite{OHD}, the cluster X-ray gas mass fraction
\cite{ref:07060033}, the measurement results of baryon acoustic
oscillation (BAO) from Sloan Digital Sky Survey (SDSS) and Two
Degree Field Galaxy Redshift Survey (2dFGRS)
\cite{SDSS}\cite{ref:Percival2}, and the current cosmic microwave
background (CMB) data from seven-year WMAP \cite{7ywmap}.

\section{$\text{Modified Chaplygin gas model}$}

 We briefly introduce the GCG model as the unification of dark matter and dark energy at first.
   The energy density $\rho$ and pressure
$p$ in this model are related by the equation of state (EOS)
\begin{equation}
p_{GCG}=-\frac{A}{\rho_{GCG}^{\alpha}},\label{GCG}
\end{equation}
where $A$  and $\alpha$ are parameters in the model. When $\alpha =
1$, it is reduced to the CG scenario. Using Eq. (\ref{GCG}) one has
a solution of energy density for the GCG fluid
\begin{equation}
\rho_{GCG}=\rho _{0GCG}[A_s+\frac{1-A_s}{a^{3(1+\alpha )}}]^{\frac
1{1+\alpha }},\label{rohGCG}
\end{equation}
where A$_s=\frac A{\rho _{0GCG}^{1+\alpha }}$. From Eq.
(\ref{rohGCG}) one can see that
 the GCG fluid behaves as a dust-like matter at early time and as  a dynamical cosmological constant
  at late epoch \cite{unification-GCG}, then the GCG  can  be
interpreted as an entangled mixture of dark matter and dark energy.
The dual role of the  GCG fluid is at the heart of the  interesting
property of this model.

 For MCG model, from phenomenological view point it is
interesting and can be motivated by the brane world interpretation
\cite{MCG-brane}. It is characterized by a more general EOS,
\begin{equation}
p_{MCG}=B\rho_{MCG} -\frac A{\rho_{MCG} ^\alpha },\label{MCG}
\end{equation}
 which
looks like that of two fluids, one obeying a perfect EOS $p=B\rho$
and the other being the GCG \cite{MCG}. Where $A, B,$ and $\alpha$
are parameters in the model, $\rho_{MCG}$ and $p_{MCG}$ are energy
density and pressure of the MCG fluid.
 From Eq. (\ref{MCG}), it is easy to see that the EOS of MCG reduces to the GCG
scenario if $B = 0$, and reduces to the perfect fluid if $A = 0$. In
addition, with $B = 0$ and $\alpha=0$ it reduces to $\Lambda$CDM
model.
 In the
space-time geometry described by the non-flat
Friedmann-Robertson-Walker (FRW) metric
\begin{equation}
ds^{2}=-dt^{2}+a^{2}(t)[\frac{dr^{2}}{1-kr^{2}}+r^{2}(d\theta^{2}+\sin^{2}
\theta d\phi^{2})],
\end{equation}
 the EOS (\ref{MCG}) leads, after
inserted into the relativistic energy conservation equation, to an
energy density  of MCG  evolving as
\begin{equation}
 \rho _{MCG}=\rho _{0MCG}[B_s+\frac{1-B_s}{a^{3(1+B)(1+\alpha)}}]^{\frac 1{1+\alpha }},\label{mcg-rho}
\end{equation}
for B $\neq $ -1, where B$_s=\frac A{(1+B)\rho _{0MCG}^{1+\alpha
}}$, $a$ is the scale factor of universe  which is related to the
redshift by, $a=\frac{1}{1+z}$. Considering that MCG fluid  plays
the role of  a mixture of  dark matter and dark energy, and assuming
that the universe is filled with three components: the  MCG, the
baryon matter, and the radiation component, one can express the
dimensionless Hubble parameter $E$ as
\begin{equation}
E=\frac{H}{H_0}=\sqrt{(1-\Omega _{b}-\Omega
_{r}-\Omega_{k})[B_s+(1-B_s)a^{-3(1+B)(1+\alpha )}]^{\frac
1{1+\alpha }}+\Omega
_{b}a^{-3}+\Omega_{r}a^{-4}+\Omega_{k}a^{-2}},\label{mcg-H}
\end{equation}
where $H$ is the  Hubble parameter, with its current value
$H_{0}=100h$ km s$^{-1}$Mpc$^{-1}$, $\Omega_{b}$, $\Omega_{r}$, and
$\Omega_{k}$ denote the dimensionless baryon matter, radiation, and
curvature density, respectively.

\section{$\text{Constraint result and conclusion}$}

 Next we apply the current observed data to constrain the MCG model. The constraint method
and the used data: Union2 SNIa,
 OHD, CBF,  BAO, and CMB data   are presented in Appendix A. In our calculations
the total likelihood function is written as $L\propto
e^{-\chi^2/2}$, with the total $\chi^2$ equaling
\begin{eqnarray}
\chi^2=\widetilde{\chi}^{2}_{SNIa}+\chi^2_{OHD}+\chi^2_{CBF}+\chi^2_{BAO}+\chi^2_{CMB},\label{chi2total}
\end{eqnarray}
where the separate likelihoods of SNIa, OHD, CBF, BAO and CMB are
given by Eqs (\ref{chi2SNABC}), (\ref{chi2OHD}),
(\ref{eq:chi2fgas}), (\ref{eq:chi2BAO}) and (\ref{eq:chi2CMB}).
From the expressions of $\chi^2_{CBF}$, $\chi^2_{BAO}$ and
$\chi^2_{CMB}$, one can see that they are related with the matter
density $\Omega_{m}$.
 For MCG model, since it is considered as the
unification of dark matter and dark energy, we do not have dark
matter in this model. So, the matter density is not explicitly
included in the background equation (\ref{mcg-H}).
 According the Eq.
(\ref{mcg-H}), by considering the universe is dominated by the
matter component  at early time ($a\ll1$), i.e., relative to the
dark matter density the dark energy density is neglectable, one can
get an effective expression of the current matter density,
$\Omega_{m}=\Omega_{b}+(1-\Omega_{b}-\Omega_{r}-\Omega_{k})(1-A_{s})^{\frac{1}{1+\alpha}}$.
This expression of $\Omega_{m}$, is an estimate of the "matter"
component of the MCG fluid with the baryon density. Thus, in the
CBF, BAO and CMB constraints, we take this expression of
$\Omega_{m}$.

Next, we perform a global fitting on determining the MCG model
parameters using the Markov Chain Monte Carlo (MCMC) approach.
 In our joint
analysis, the MCMC code is based on the publicly available
\textbf{CosmoMC} package \cite{ref:MCMC} and the  \textbf{modified
CosmoMC} package \cite{ref:0409574,ref:07060033,ref:modifiedMCMC}.
The latter package is about the constraint code  of X-ray cluster
gas mass fraction, with including additional 7 free parameters
$(K,\eta,\gamma,b_{0},\alpha_{b},s_{0},\alpha_{s})$. In these
packages, they have been modified to include the new model
parameters $B_{s}$, $\alpha$ and B. In the calculation the baryon
matter density is taken to be varied with a tophat prior:
$\Omega_{b}h^{2}\in [0.005,0.1]$. In addition, for the MCMC
calculation on the MCG model, we run 8 independent chains, and to
get the converged results we test the convergence of the chains by
typically getting $R - 1$ to be less than 0.03.

In FIG. \ref{fig:contoursMCG-flat},  we show a one-dimensional
probability  distribution of each parameter and two-dimensional
plots for parameters between each other  in the flat MCG model.
According to the figure, the constraint results on the best fit
values of parameters with $1\sigma$ and $2\sigma$ confidence levels
are listed in Table \ref{tab:resultsMCG}. In Ref. \cite{MCG-sound}
it is shown that for the constraint on MCG model parameters obtained
from the sound speed, $0<c_{s}^{2}<c^{2}$, the parameters are
restricted to $0<(B+1)(\alpha+1)<2$. According to this constraint on
MCG model parameters one can see that the MCG fluid has a well sound
speed. Considering Ref. \cite{MCG-constraint}, where the model
parameter $B$ is constrained by using the location of the peak of
the CMB spectrum, $-0.35\lesssim B \lesssim 0.025$, it is easy to
see that our result is more stringent. In addition, considering Ref.
\cite{MCG-spectrum}, where the values of model parameters are
analyzed against the matter power spectrum observational data, it is
obtained that a very stringent constraint exists on $B$, which is
consistent with our result, i.e., it seems that the MCG model is
viable only for very special cases, and it tends to reduce to the
GCG scenario.

\begin{figure}[!htbp]
  \includegraphics[width=18cm]{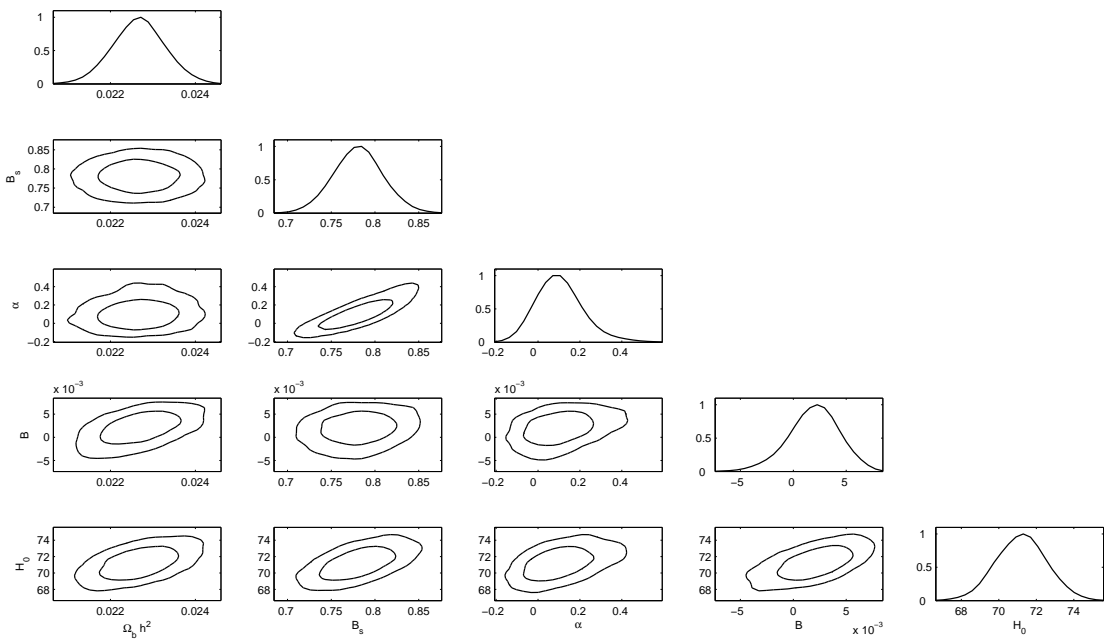}\\
 \caption{The 2-D contours with  $1\sigma, 2\sigma$
  confidence levels and 1-D marginalized
  distribution
  of  $\Omega_{b}h^2$, $B_{s}$, $\alpha$, $B$, and $H_0$ in the flat  MCG model.}\label{fig:contoursMCG-flat}
\end{figure}

\begin{table}
\begin{center}
\begin{tabular}{cc    }
 \hline\hline Parameters &  Best fit values                                        \\ \hline
 $\Omega_{b}h^2$  &  $0.02263^{+0.00184 +0.00213}_{-0.00162 -0.00195}$    \\
 $B_{s}$          &  $0.7788^{+0.0736 +0.0918}_{-0.0723 -0.0904}$          \\
 $\alpha$         &  $0.1079^{+0.3397 +0.4678}_{-0.2539 -0.2911}$         \\
  $B$               &  $0.00189^{+0.00583 +0.00660}_{-0.00756 -0.00915}$     \\
 $H_0$            &  $70.711^{+4.188 +5.281}_{-3.142 -4.149}$               \\  \hline
 $\chi^2_{min}(\chi^{2}_{min}/dof)$    & 596.147 (0.9709)                   \\
\hline\hline
\end{tabular}
\caption{The data fitting results of the MCG  model parameters with
$1\sigma$ and $2\sigma$ confidence levels.}\label{tab:resultsMCG}
\end{center}
\end{table}

 In addition,  we consider a more general non-flat background geometry.
For using  above combined observational data to constrain the  MCG
model, since the space curvature $\Omega_{k}$ is near to zero, the
constraint results of the model parameters obtained from the
non-flat universe are similar to the cases of  the flat universe.
 For simplicity, we do not list the constraint results for this case. With replacing the Union2 SNIa data\footnote{The
Union2 SNIa data are obtained, by  adding  new datapoints (including
the high redshift SNIa) to the Union SNIa data, making a number of
refinements to the Union analysis chain, refitting all light curves
with the SALT2 fitter.} with the 397 Constitution data\footnote{The
397 Constitution data are obtained by adding 90 SNIa from CfA3
sample to 307 SNIa Union sample. CfA3 sample are all from the
low-redshift SNIa, $z<0.08$, and these 90 SNIa are calculated with
using the same Union cuts.}
 \cite{397Constitution} in the above combined constraint, we obtain the constraint
 results  of  MCG model parameters,
  $\Omega_{k}=-0.000844_{-0.015191}^{+0.013471}$ $(1\sigma)$ $_{-0.017862}^{+0.014536}$ $(2\sigma)$,
  $B_{s}=0.7541_{-0.0892}^{+0.0941}$ $(1\sigma)$ $_{-0.0965}^{+0.1092}$ $(2\sigma)$,
  $\alpha=0.0082_{-0.2318}^{+0.3837}$ $(1\sigma)$ $_{-0.2433}^{+0.4401}$ $(2\sigma)$,
  $B=0.00138_{-0.00738}^{+0.00817}$ $(1\sigma)$ $_{-0.00869}^{+0.00931}$
  $(2\sigma)$, with $\chi^{2}_{min}=520.118$ ($\chi^{2}_{min}/dof=1.148$). It seems that for this case, the
  MCG model tends to reduce to the flat cosmic concordance model, $\Lambda$CDM.

 \textbf{\ Acknowledgments }
 The research work is supported by the National Natural Science Foundation
  (Grant No. 10875056), NSF (10703001) and NSF (No.11005088)  of P.R.
  China.

\appendix
\section{$\text{Current observational data and cosmological constraints}$}\label{constraint-method}
 In this part we introduce the cosmological constraint methods and the current observed data used in this paper.
\subsection{Type Ia supernovae}

 SNIa behave as the excellent standard
 candles, so they can be used to directly measure the expansion rate of
 the universe from the high redshift to the present
 time. For using SNIa data, theoretical dark-energy model parameters are
determined by minimizing the quantity \cite{chi2SNe}
\begin{equation}
\chi^{2}_{SNIa}(\mu_{0},
p_{s})=\sum_{i=1}^{N}\frac{(\mu_{obs}(z_{i})
-\mu_{th}(z_{i};\mu_{0},p_{s}))^2}{\sigma^2_{\mu_{obs}}(z_{i})},\label{chi2SN}
\end{equation}
where $N=557$ for Union2 dataset, which is the largest  SNIa sample
by far;  $p_{s}$ denotes the model parameters;
$\sigma_{\mu_{obs}}(z_{i})$ are errors; $\mu_{obs}(z_{i})=
m_{obs}(z_{i})-M$, is the observed value of distance modulus of SNIa
at $z_{i}$ and can be given by the SNIa dataset; $\mu_{th}$ is the
theoretical distance modulus, which is related to the apparent
magnitude of
  SNIa at peak brightness $m$ and the absolute magnitude $M$,
\begin{equation}
\mu_{th}(z)\equiv
m_{th}(z)-M=5log_{10}(D_{L}(z))+\mu_{0}.\label{muth}
\end{equation}
Here, the Hubble free luminosity distance
\begin{equation}
D_{L}(z)=H_{0}d_{L}(z)=\frac{c(1+z)}{\sqrt{| \Omega_{k}|}}
sinn[\sqrt{| \Omega_{k}|}
\int_{0}^{z}\frac{H_{0}dz^{'}}{H(z^{'};p_{s})}],\label{dl}
\end{equation}
and
\begin{equation}
\mu_{0}=5log_{10}(\frac{H_{0}^{-1}}{Mpc})+25=42.38-5log_{10}h,\label{mu0}
\end{equation}
 with $h$ being  a re-normalized quantity, which is given by
 $H_{0}=100h$ km s$^{-1}$Mpc$^{-1}$. It should be noted that $\mu_{0}$ is independent of the data and the
dataset. By expanding the $\chi^{2}$ of Eq. (\ref{chi2SN}) relative
to the nuisance parameter $\mu_{0}$,  the minimization with respect
to $\mu_{0}$ can be made trivially \cite{chi2SNeABC}
\begin{equation}
\chi^{2}_{SNIa}(p_{s})=A(p_{s})-2\mu_{0}B(p_{s})+\mu_{0}^{2}C,\label{chi2SN1}
\end{equation}
where
\begin{equation}
A(p_{s})=\sum_{i=1}^{N}\frac{[\mu_{obs}(z_{i})-\mu_{th}(z_{i};\mu_{0}=0,p_{s})]^{2}}{\sigma^2_{\mu_{obs}}(z_{i})},\label{chi2SNA}
\end{equation}
\begin{equation}
B(p_{s})=\sum_{i=1}^{N}\frac{\mu_{obs}(z_{i})-\mu_{th}(z_{i};\mu_{0}=0,p_{s})}{\sigma^2_{\mu_{obs}}(z_{i})},\label{chi2SNB}
\end{equation}
\begin{equation}
C=\sum_{i=1}^{N}\frac{1}{\sigma^2_{\mu_{obs}}(z_{i})}.\label{chi2SNC}
\end{equation}
Obviously, according to Eq. (\ref{chi2SN1}) $\chi^{2}_{SNIa}$ has a
minimum for $\mu_{0}=B/C$. Thus,  the expression of $\chi^{2}$ for
SNIa constraint can be written as
\begin{equation}
\widetilde{\chi}^{2}_{SNIa}(p_{s})=A(p_{s})-B(p_{s})^{2}/C.\label{chi2SNABC}
\end{equation}
 Since $\chi^{2}_{SNIa,min}=\widetilde{\chi}^{2}_{SNIa,min}$ and $\widetilde{\chi}^{2}_{SNIa}$ is independent of the nuisance
parameter $\mu_{0}$, one usually utilize
 the expression (\ref{chi2SNABC}) to displace (\ref{chi2SN}) to perform the likelihood analysis for the SNIa
constraint. For minimizing $\chi^{2}_{SNIa}(p_{s},B/C)$ to constrain
cosmological model, it is equivalent to maximizing the likelihood
\begin{equation}
L(p_{s})\propto \exp[\frac{-\chi^{2}(p_{s})}{2}].\label{Lsn}
\end{equation}

\subsection{Observational Hubble data}

The observational Hubble data \cite{ohdzhang} are based on
differential ages of the galaxies. In \cite{ref:JVS2003}, Jimenez
{\it et al.} obtained an independent estimate for the Hubble
parameter using the method developed in \cite{ref:JL2002}, and used
it to constrain the cosmological models. The Hubble parameter
depending on the differential ages as a function of redshift $z$ can
be written in the form of
\begin{equation}
H(z)=-\frac{1}{1+z}\frac{dz}{dt}.
\end{equation}
So, once $dz/dt$ is known, $H(z)$ is obtained directly. By using the
differential ages of passively-evolving galaxies from the Gemini
Deep Deep Survey (GDDS) \cite{ref:GDDS} and archival data
\cite{ref:archive1,ref:archive2,ref:archive3,ref:archive4,ref:archive5,ref:archive6},
Simon {\it et al.} obtained several values of $H(z)$  at  different
redshift \cite{OHD}. The twelve observational Hubble data  (redshift
interval  $0\lesssim z \lesssim 1.8$) from Refs.
\cite{12Hubbledata,H0prior} are listed in Table
\ref{table-12Hubbledata}.
\begin{table}[ht]
\begin{center}
\begin{tabular}{c|llllllllllll}
\hline\hline
 $z$ &\ 0 & 0.1 & 0.17 & 0.27 & 0.4 & 0.48 & 0.88 & 0.9 & 1.30 & 1.43 & 1.53 & 1.75  \\ \hline
 $H(z)\ ({\rm km~s^{-1}\,Mpc^{-1})}$ &\ 74.2 & 69 & 83 & 77 & 95 & 97 & 90 & 117 & 168 & 177 & 140 & 202  \\ \hline
 $1 \sigma$ uncertainty &\ $\pm 3.6$ & $\pm 12$ & $\pm 8$ & $\pm 14$ & $\pm 17$ & $\pm 60$ & $\pm 40$
 & $\pm 23$ & $\pm 17$ & $\pm 18$ & $\pm 14$ & $\pm 40$ \\
\hline\hline
\end{tabular}
\end{center}
\caption{\label{table-12Hubbledata} The observational $H(z)$
data~\cite{12Hubbledata,H0prior}.}
\end{table}
In addition, in \cite{3Hubbledata} the authors take the BAO scale as
a standard ruler in the radial direction, and obtain three
additional data: $H(z=0.24)=79.69\pm2.32, H(z=0.34)=83.8\pm2.96,$
and $H(z=0.43)=86.45\pm3.27$.

 The best fit values of the model parameters from observational Hubble data  are determined by minimizing \cite{chi2hub}
 \begin{equation}
 \chi_{OHD}^2(H_{0},p_{s})=\sum_{i=1}^{15} \frac{[H_{th}(H_{0},p_{s};z_i)-H_{obs}(z_i)]^2}{\sigma^2(z_i)},\label{chi2OHD}
 \end{equation}
 where  $H_{th}$ is the predicted value of the Hubble parameter, $H_{obs}$ is the observed value, $\sigma(z_i)$ is the standard
 deviation measurement uncertainty, and the summation is over the $15$ observational Hubble data points at redshifts $z_i$.

\subsection{The X-ray gas mass fraction}
The X-ray gas mass fraction, $f_{gas}$, is defined as the ratio of
the X-ray gas mass to the total mass of a cluster, which is
approximately independent on the redshift for the hot ($kT \gtrsim
5keV$), dynamically relaxed clusters at the radii larger than the
innermost core $r_{2500}$. As inspected in \cite{ref:07060033}, the
$\Lambda$CDM model is very favored and has been chosen as the
reference cosmology. The model fitted to the reference $\Lambda$CDM
data is presented as \cite{ref:07060033}
\begin{eqnarray}
&&f_{gas}^{\Lambda CDM}(z)=\frac{K A \gamma
b(z)}{1+s(z)}\left(\frac{\Omega_b}{\Omega_m}\right)
\left[\frac{D_A^{\Lambda CDM}(z)}{D_A(z)}\right]^{1.5},\ \ \ \
\label{eq:fLCDM}
\end{eqnarray}
 where $D_{A}^{\Lambda CDM} (z)$ and $D_{A}(z)$ denote respectively
the proper angular diameter distance in the $\Lambda$CDM reference
cosmology and the current constraint model. $A$ is the angular
correction factor, which is caused by the change in angle for the
current test model $\theta_{2500}$ in comparison with that of the
reference cosmology $\theta_{2500}^{\Lambda CDM}$:
\begin{eqnarray}
&&A=\left(\frac{\theta_{2500}^{\Lambda
CDM}}{\theta_{2500}}\right)^\eta \approx
\left(\frac{H(z)D_A(z)}{[H(z)D_A(z)]^{\Lambda CDM}}\right)^\eta,
\end{eqnarray}
here, the index $\eta$ is the slope of the $f_{gas}(r/r_{2500})$
data within the radius $r_{2500}$, with the best-fit average value
$\eta=0.214\pm0.022$ \cite{ref:07060033}. And the proper (not
comoving) angular diameter distance is given by
\begin{eqnarray}
&&D_A(z)=\frac{c}{(1+z)\sqrt{|\Omega_k|}}\mathrm{sinn}[\sqrt{|\Omega_k|}\int_0^z\frac{dz'}{H(z';p_{s})}],
\end{eqnarray}
which is related with $d_{L}(z)$ by
\begin{equation}
D_A(z)=\frac{d_{L}(z)}{(1+z)^2},\nonumber
\end{equation}
where $sinn(\sqrt{|\Omega_k|}x)$, respectively, denotes
$\sin(\sqrt{|\Omega_k|}x)$, $\sqrt{|\Omega_k|}x$,
$\sinh(\sqrt{|\Omega_k|}x)$ for $\Omega_k<0$, $\Omega_k=0$ and
$\Omega_k>0$.

In equation (\ref{eq:fLCDM}), the parameter $\gamma$ denotes
permissible departures from the assumption of hydrostatic
equilibrium, due to non-thermal pressure support; the bias factor
$b(z)= b_0(1+\alpha_b z)$ accounts for uncertainties in the cluster
depletion factor; $s(z)=s_0(1 +\alpha_s z)$ accounts for
uncertainties of the baryonic mass fraction in stars and a Gaussian
prior for $s_0$ is employed, with $s_0=(0.16\pm0.05)h_{70}^{0.5}$
\cite{ref:07060033}; the factor $K$ is used to describe the combined
effects of the residual uncertainties, such as the instrumental
calibration and certain X-ray modelling issues, and a Gaussian prior
for the 'calibration' factor is considered by $K=1.0\pm0.1$
\cite{ref:07060033}.

Following the method in Ref. \cite{ref:CBFchi21,ref:07060033} and
adopting the updated 42 observational $f_{gas}$ data in Ref.
\cite{ref:07060033}, the best fit values of the model parameters for
the X-ray gas mass fraction analysis are determined by minimizing,
\begin{eqnarray}
&&\chi^2_{CBF}=\sum_i^N\frac{[f_{gas}^{\Lambda
CDM}(z_i)-f_{gas}(z_i)]^2}{\sigma_{f_{gas}}^2(z_i)}+\frac{(s_{0}-0.16)^{2}}{0.0016^{2}}
+\frac{(K-1.0)^{2}}{0.01^{2}}+\frac{(\eta-0.214)^{2}}{0.022^{2}},\label{eq:chi2fgas}
\end{eqnarray}
where $\sigma_{f_{gas}}(z_{i})$ is the statistical uncertainties
(Table 3 of \cite{ref:07060033}). As pointed out in
\cite{ref:07060033}, the acquiescent systematic uncertainties have
been considered according to the parameters i.e. $\eta$, $b(z)$,
$s(z)$ and $K$.

\subsection{Baryon acoustic oscillation}

The baryon acoustic oscillations are detected in the clustering of
the combined 2dFGRS  and   SDSS main galaxy samples, which measure
the
 distance-redshift relation at $z_{BAO} = 0.2$ and $z_{BAO} = 0.35$. The observed scale of the BAO calculated from these samples, are
 analyzed using estimates of the correlated errors to constrain the form of the distance measure $D_V(z)$  \cite{ref:Okumura2007,ref:Percival2}
\begin{equation}
D_V(z)=\left[(1+z)^2 D^2_A(z) \frac{cz}{H(z)}\right]^{1/3}.
\label{eq:DV}
\end{equation}
The peak positions of the BAO depend on the ratio of $D_V(z)$ to the
sound horizon size at the drag epoch (where baryons were released
from photons) $z_d$, which can be obtained by using a fitting
formula \cite{27Eisenstein}:
\begin{eqnarray}
&&z_d=\frac{1291(\Omega_mh^2)^{-0.419}}{1+0.659(\Omega_mh^2)^{0.828}}[1+b_1(\Omega_bh^2)^{b_2}],
\end{eqnarray}
with
\begin{eqnarray}
&&b_1=0.313(\Omega_mh^2)^{-0.419}[1+0.607(\Omega_mh^2)^{0.674}], \\
&&b_2=0.238(\Omega_mh^2)^{0.223}.
\end{eqnarray}
In this paper, we use the data of $r_s(z_d)/D_V(z)$ extracted from
the Sloan Digitial Sky Survey (SDSS) and the Two Degree Field Galaxy
Redshift Survey (2dFGRS) \cite{ref:Okumura2007}, which are listed in
Table \ref{baodata}, where $r_s(z)$ is the comoving sound horizon
size
\begin{eqnarray}
r_s(z)&&{=}c\int_0^t\frac{c_sdt}{a}=c\int_0^a\frac{c_sda}{a^2H}=c\int_z^\infty
dz\frac{c_s}{H(z)} \nonumber\\
&&{=}\frac{c}{\sqrt{3}}\int_{0}^{1/(1+z)}\frac{da}{a^2H(a)\sqrt{1+(3\Omega_b/(4\Omega_\gamma)a)}},
\end{eqnarray}
where $c_s$ is the sound speed of the photon$-$baryon fluid
\cite{ref:Hu1, ref:Hu2}:
\begin{eqnarray}
&&c_s^{-2}=3+\frac{4}{3}\times\frac{\rho_b(z)}{\rho_\gamma(z)}=3+\frac{4}{3}\times(\frac{\Omega_b}{\Omega_\gamma})a,
\end{eqnarray}
and here $\Omega_\gamma=2.469\times10^{-5}h^{-2}$ for
$T_{CMB}=2.725K$.

\begin{table}[htbp]
\begin{center}
\begin{tabular}{c|l}
\hline\hline
 $z$ &\ $r_s(z_d)/D_V(z)$  \\ \hline
 $0.2$ &\ $0.1905\pm0.0061$  \\ \hline
 $0.35$  &\ $0.1097\pm0.0036$  \\
\hline
\end{tabular}
\end{center}
\caption{\label{baodata} The observational $r_s(z_d)/D_V(z)$
data~\cite{ref:Percival2}.}
\end{table}
Using the data of BAO in Table \ref{baodata} and the inverse
covariance matrix $V^{-1}$ in \cite{ref:Percival2}:
\begin{eqnarray}
&&V^{-1}= \left(
\begin{array}{cc}
 30124.1 & -17226.9 \\
 -17226.9 & 86976.6
\end{array}
\right),
\end{eqnarray}
  the $\chi^2_{BAO}(p_s)$ is given as
\begin{equation}
\chi^2_{BAO}(p_s)=X^tV^{-1}X,\label{eq:chi2BAO}
\end{equation}
where $X$ is a column vector formed from the values of theory minus
the corresponding observational data, with
\begin{eqnarray}
&&X= \left(
\begin{array}{c}
 \frac{r_s(z_d)}{D_V(0.2)}-0.1905 \\
 \frac{r_s(z_d)}{D_V(0.35)}-0.1097
\end{array}
\right),
\end{eqnarray}
and $X^t$ denotes its transpose.

\subsection{Cosmic microwave background}

The CMB shift parameter $R$ is provided by \cite{ref:Bond1997}
\begin{equation}
R(z_{\ast})=\sqrt{\Omega_m H^2_0}(1+z_{\ast})D_A(z_{\ast})/c,
\end{equation}
here, the redshift $z_{\ast}$ (the decoupling epoch of photons) is
obtained using the fitting function \cite{Hu:1995uz}
\begin{equation}
z_{\ast}=1048\left[1+0.00124(\Omega_bh^2)^{-0.738}\right]\left[1+g_1(\Omega_m
h^2)^{g_2}\right],\nonumber
\end{equation}
where the functions $g_1$ and $g_2$ read
\begin{eqnarray}
g_1&=&0.0783(\Omega_bh^2)^{-0.238}\left(1+ 39.5(\Omega_bh^2)^{0.763}\right)^{-1},\nonumber \\
g_2&=&0.560\left(1+ 21.1(\Omega_bh^2)^{1.81}\right)^{-1}.\nonumber
\end{eqnarray}
In addition, the acoustic scale is related to a distance ratio,
$D_A(z)/r_s(z)$,   and at decoupling epoch it is defined as
\begin{eqnarray}
&&l_A\equiv(1+z_{\ast})\frac{\pi
D_A(z_{\ast})}{r_s(z_{\ast})},\label{la}
\end{eqnarray}
where Eq.(\ref{la}) arises a factor  $1+z_{\ast}$, because $D_A(z)$
is the proper (physical) angular diameter distance, whereas
$r_{s}(z_{\ast})$ is the comoving sound horizon. Using the data of
$l_A, R, z_\ast$ in \cite{7ywmap} and their covariance matrix of
$[l_A(z_\ast), R(z_\ast), z_\ast]$ (please see table
\ref{tab:7yearWMAPdata} and \ref{tab:7yearWMAPcovariance}), we can
calculate the likelihood $L$ as $\chi^2_{CMB}=-2\ln L$:
\begin{eqnarray}
&&\chi^2_{CMB}=\bigtriangleup d_i[Cov^{-1}(d_i,d_j)[\bigtriangleup
d_i]^t],\label{eq:chi2CMB}
\end{eqnarray}
where $\bigtriangleup d_i=d_i-d_i^{data}$ is a row vector, and
$d_i=(l_A, R, z_\ast)$.\\

 \begin{table}
 \begin{center}
 \begin{tabular}{c c   cc   } \hline\hline
 ~ &              7-year maximum likelihood ~~~ & error, $\sigma$ &\\ \hline
 $ l_{A}(z_{\ast})$         & 302.09      & 0.76  & \\
 $ R(z_{\ast})$             &  1.725      & 0.018 & \\
 $ z_{\ast}$                & 1091.3     & 0.91&    \\
 \hline\hline
 \end{tabular}
 \caption{The values of  $ l_{A}(z_{\ast})$, $R(z_{\ast})$, and $z_{\ast}$ from 7-year WMAP data.}\label{tab:7yearWMAPdata}
 \end{center}
 \end{table}

\begin{table}
 \begin{center}
 \begin{tabular}{c c   cc c  } \hline\hline
 ~ &             $ l_{A}(z_{\ast})$      & $ R(z_{\ast})$   & $ z_{\ast}$ &  \\ \hline
 $ l_{A}(z_{\ast})$         & 2.305      & 29.698           &  -1.333     & \\
 $ R(z_{\ast})$             &  ~         & 6825.270         &  -113.180    & \\
 $ z_{\ast}$                & ~          & ~                &  3.414      &  \\
 \hline\hline
 \end{tabular}
 \caption{The inverse covariance matrix of  $ l_{A}(z_{\ast})$, $R(z_{\ast})$, and $z_{\ast}$ from 7-year WMAP data.}\label{tab:7yearWMAPcovariance}
 \end{center}
 \end{table}

\end{document}